\documentclass[%
 aip,
 amsmath,amssymb,
 reprint,%
]{revtex4-1}

\usepackage{graphicx}% Include figure files
\usepackage{dcolumn}% Align table columns on decimal point
\usepackage{bm}% bold math
\usepackage[normalem]{ulem}

\usepackage[utf8]{inputenc}
\usepackage[T1]{fontenc}
\usepackage{mathptmx}
\usepackage{etoolbox}
\usepackage{xcolor}
\usepackage{hyperref}

\newcommand{\sm}{\ensuremath{\mathbf{S}}}

\makeatletter
\def\@email#1#2{%
 \endgroup
 \patchcmd{\titleblock@produce}
  {\frontmatter@RRAPformat}
  {\frontmatter@RRAPformat{\produce@RRAP{*#1\href{mailto:#2}{#2}}}\frontmatter@RRAPformat}
  {}{}
}%
\makeatother
\begin{document}

\preprint{AIP/123-QED}

%\title{Conditions of all-optical switching of GdCo alloy in ultra-thin film and XML\\
\title{Atomistic Insights into the All-Optical Switching of GdCo Alloys with Minimal Gd Concentration}
% Force line breaks with \\
\author{Paul-Iulian Gavriloaea}
\affiliation{National Institute of Research and Development for Technical Physics, 47 Mangeron Blvd., 700050 Iasi, Romania}
\affiliation{Instituto de Ciencia de Materiales de Madrid, CSIC, Cantoblanco, 28049 Madrid, Spain}
%Lines break automatically or can be forced with \\
\author{Jackson L. Ross}%
%\affiliation{ <Elton prefers I use Edinburgh only for his projects>
%Institute for Condensed Matter Physics and Complex Systems, The
%University of Edinburgh, Edinburgh, EH9 3FD, United Kingdom}%
\affiliation{School of Physics, Engineering and Technology, University of York, YO10 5DD, York, UK}
\author{Richard F. L. Evans}
\affiliation{School of Physics, Engineering and Technology, University of York, YO10 5DD, York, UK}
\author{Roy Chantrell}
\affiliation{School of Physics, Engineering and Technology, University of York, YO10 5DD, York, UK}
\author{Stéphane Mangin}
\affiliation{Université de Lorraine, CNRS, IJL, Nancy, France}
\author{Oksana Chubykalo-Fesenko}
\affiliation{Instituto de Ciencia de Materiales de Madrid, CSIC, Cantoblanco, 28049 Madrid, Spain}

\date{\today}% It is always \today, today,
             %  but any date may be explicitly specified

\begin{abstract}
Single-pulse all-optical helicity-independent switching (AO-HIS) is most commonly observed in Gd-based ferrimagnets near magnetization compensation. Recent experiments, however, demonstrated switching in Pt/Co/Pt-based heterostructures containing only a small amount of Gd at a Co/Pt interface. Here, we use atomistic spin-dynamics simulations to test an effective interfacial GdCo-alloy picture and to determine how species-dependent spin-bath coupling controls the minimum Gd concentration required for reversal. We compare a three-monolayer Co film with Gd confined to the surface layer and a three-dimensional uniformly alloyed GdCo model. Within the present uniform-heating model, robust transient toggle-like reversal is obtained far from compensation when the Gd coupling to the thermal bath is substantially weaker than that of Co; for $\lambda_{\text{Co}} = 0.02$ and $\lambda_{\text{Gd}} = 0.001$, reversal occurs down to 1.5 at.\% total Gd for selected pulse parameters. Increasing the Gd content accelerates reversal, consistent with the interpretation of Gd as an angular-momentum reservoir, whereas dilute interfacial Gd produces slower and more stochastic dynamics. These results show that compensation is not a fundamental requirement for ultrafast AO-HIS and identify the effective Gd spin-bath coupling as a key control parameter. Quantitative comparison with experiment will require explicit optical absorption, stack-dependent heat flow, dipolar fields, and a stable-final-state switching criterion.
\end{abstract}

\maketitle

\section{\label{sec:level1}Introduction}
All-optical helicity independent switching (AO-HIS) has attracted significant attention due to its potential to enable ultrafast control of magnetization without the need for external magnetic fields.\cite{Stanciu2007,Radu2011TransientSpins,Ostler2012UltrafastFerrimagnet} This capability makes AO-HIS a promising mechanism for next-generation data storage and spintronic devices, where speed, scalability, and reduced energy consumption are critical.\cite{Kimel2019WritingPulses,REMYandManging2026} In particular, the ability to manipulate magnetic states on femtosecond timescales offers a pathway toward technological advancements beyond the limits of conventional magnetic recording.
\newline
\indent
AO-HIS has been predominantly observed in Gd-based transition metal rare-earth alloys and multilayers such as GdFeCo,\cite{Stanciu2007,Radu2011TransientSpins,Ostler2012UltrafastFerrimagnet} GdCo,\cite{GdCo_AO_HIS} GdTbCo,\cite{Ceballos2021,ZhangGdStateDiagram} or Gd/Co.\cite{Lalieu2017,Kboonthum_controlling_single_pulse} More recently, it was also confirmed in two classes of Gd-free samples: the Heusler alloy MnRuGa \cite{Banerjee2020} and several ferrimagnetic multilayers such as Tb/Co,\cite{LAviles2019,LAviles2020} Dy/Co or Tb/Fe \cite{Peng2023} and Co/Ho. \cite{CoHo_AO_HIS_2023} 
% Among these however, GdFeCo and GdCo stand out as prototypical systems for investigating AO-HIS, owing to its robustness and the extensive experimental validation of its switching behavior under a wide range of conditions.
Regarding the necessary ingredients for AO-HIS to occur and restricting ourselves to Gd-based systems, the role of the magnetization compensation temperature $T_{\text{M}}$ has long been put into question. This constitutes an ongoing debate as numerous experimental studies on perpendicularly magnetized GdFeCo or GdCo alloys demonstrated AO-HIS can be achieved both above and below $T_\text{M}$, \cite{Ostler2012UltrafastFerrimagnet,LeGuyader2016,Xu2017,Jakobs2021,Wei2021,Zhang2024} yet switching remained primarily limited to alloy compositions in the vicinity of this compensation point. Distinctively, for GdCo alloys with in-plane magnetisation, AO-HIS is obtained in a much wider Gd concentration range.\cite{Lin2023} More interestingly, for layered synthetic ferrimagnets such as Gd/Co characterised also by perpendicular magnetization geometry, AO-HIS appears independent of $T_\text{M}$ although one can argue magnetic compensation arises at the bilayer interface. \cite{Lalieu2017,Beens_alloy_syntethic_ML} 
% , several theoretical works demonstrated that AO-HIS is most energy-efficient when the initial sample temperature is in the vicinity of the magnetization compensation temperature $T_{\text{M}}$.\cite{closetocomp1,closetocomp2,closetocomp3} 
% This experimental trend was supported by 
In support of these experimental observations, the atomistic simulations of Barker et al~\cite{Barker2013TwoMagnon} predicted a switching phase diagram where AO-HIS is favored near $T_{\text{M}}$.
The authors showed that a non-equilibrium energy transfer between ferromagnetic- and antiferromagnetic-like magnon branches is maximized near the compensation point. This enhanced energy transfer leads to a minimum threshold for AO-HIS in this region. Consistent with this picture, Beens et al.~\cite{Beens_alloy_syntethic_ML} used a microscopic three-temperature model that incorporates exchange scattering processes, arriving at the conclusion that AO-HIS in GdCo alloy can only be obtained in a limited concentration range around the magnetic compensation point unlike in Gd/Co bilayers where AO-HIS is independent of $T_\text{M}$. This interpretation was later refined by F. Jakobs and U. Atxitia,\cite{Unai_universal_criteria_ferrimagnets} who demonstrated that the essential requirement for switching is not the exact compensation condition itself, but rather reducing the net magnetization under a sufficiently low threshold during laser pulse heating. In this framework, alloy compositions or initial sample temperatures near the compensation point naturally satisfy this condition, explaining their prominence in experiments. 
%Their atomistic spin simulations  revealed that ferrimagnetic alloys with low net magnetization, strong temperature dependence of magnetization (dM/dT), and weak electron–phonon coupling tend to require lower energy input to achieve switching. 
%Their results yielded a similar switching phase diagram, reinforcing the idea that reduced magnetization plays a central role.
\newline\indent 
Conditions of switching in rare earth-transition metal alloys and multilayers also involve a necessity of different demagnetizing times of the two components, based on the ratio $\mu /\lambda T_\text{C}$ \cite{Koopmans2009,Atxitia2011,Atxitia2014} where $\mu$ is the atomic magnetic moment, $T_\text{C}$ is the Curie temperature and $\lambda$ is an intrinsic damping physically representing the coupling of the spin to the heat bath. \cite{QuantumLLB} While the influence of the $\mu /T_\text{C}$ ratio  has been extensively studied, the role of the intrinsic damping is frequently overlooked.
%Later Suarez et al \cite{Suarez} have discussed a more complex behavior in alloys showing that the longitudinal magnetisation relaxation timescale  is inversely proportional to $H_{ex}^{\nu}=\mu_{\nu}( x_{\nu}J_{0,\nu}m_{\nu}+(1-x_{\nu}|J_{0,\nu \kappa})|m_{\kappa})$ where $\nu,\kappa$ stand for Gd or Fe contributions   and $J_0$ for the corresponding exchange parameters, including the average nearest number of neighbors in the undoped lattice. Based on this, Fe and Gd timescale can accelerates at low temperatures and desaccelerates at high temperatures with a maximum peak positions at different temperatures for Gd and Fe
Based solely on the $\mu /T_\text{C}$ term, the rare-earth component 
\begin{figure}
\centering
\includegraphics{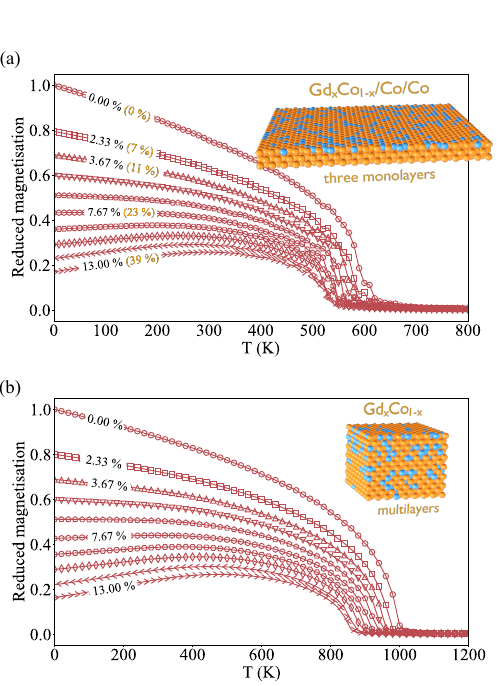}% Here is how to import EPS art
\caption{\textbf{Reduced net magnetisation as a function of temperature at distinct alloy compositions:} (a) 3ML (Gd$_{x}$Co$_{1-x}$/Co/Co) and (b) XML (Gd$_{x}$Co$_{1-x}$) system. The magnetisation is calculated via: $|{\bm{m}}|=|\sum_i\mu_\text{s}^i\sm_i|/\sum_i\mu_\text{s}^i$. The Gd concentration is reported as a percentage with respect to the total atomic volume. From $x=2.33\%$ to $x=13\%$ we display $m(T)$ curves with a concentration step of approximately $1.33\%$. For the 3ML case, we also provide in parentheses the Gd concentration calculated with respect to the atomic population of the first monolayer.  The inset of each subplot schematically displays the simulated geometry.}
\label{Figure_1}
\end{figure}
is typically four to ten times much slower compared to the transition metal \cite{Atxitia2014,Moreno2019} thereby assuring the appearance of a transient ferromagnetic-like state during AOS. \cite{Ostler2012UltrafastFerrimagnet,Atxitia2014}
% Although the microscopic Elliott-Yafet spin-flip probability $\alpha_{\text{sf}}$ in Gd is comparable to that of 3d transition metals such as Ni and Co, \cite{Koopmans2009,Steiauf2009,Steiauf2010} its slower rate of demagnetization and recovery are explained by the larger $\mu /T_\text{C}$ figure of merit. This has been understood with the aid of the well known M3TM model employed by Koopmans et al. \cite{Koopmans2009}, where the rate of the magnetization dynamics is proportional to $\alpha_{\text{sf}}T_\text{C}^2/\mu$.
% A complete picture regarding the involved spin-flip mechanisms in ultrafast magnetisation dynamics represents still a heated debate today which goes well beyond the Elliott Yafet phonon mediated scattering discussed for instance in Ref \cite{Koopmans2009}.
At the same time, tuning the magnitude of the phenomenological damping parameter $\lambda$ was shown to yield type-I or type-II magnetisation dynamics in the high-temperature micromagnetic framework of the Landau-Lifshitz-Bloch equation. U. Atxitia and O. Chubykalo-Fesenko in Ref.\cite{Atxitia2011} showed that large $\lambda$ values favor faster demagnetization and magnetization recovery associated to transition metals whereas a two orders of magnitude smaller $\lambda$ yields the slower two step demagnetization in Gd. Traditionally, in many atomistic spin dynamics (ASD) models of ferrimagnetic alloys, \cite{Radu2011TransientSpins,Barker2013TwoMagnon,Atxitia2014,Suarez2015,Moreno2017,Moreno2019,Unai_universal_criteria_ferrimagnets} the damping parameters  were considered equal in between the distinct atomic species based on the assumption of shared itinerant electrons. This hypothesis may fail at low Gd concentrations for instance, where large pure transition-metal islands are expected. Additionally, even at large Gd concentrations chemical segregation was shown to break spatial homogeneity, \cite{Graves2013} which results in a distinct local reversal at the nanometer level starting in Gd-rich regions as opposed to the well known macroscopic behaviour of GdFeCo. \cite{Radu2011TransientSpins} Finally, several other works highlighted the importance of element specific damping in capturing the physics of AO-HIS, for instance in order to model the reversal of GdFeCo in a wide fs-ps pulse range \cite{Jakobs2021} or to explain experimental switching dynamics in GdTbCo alloys. \cite{Ceballos2021}
\newline\indent
%In addition to compositional effects, the switching behavior also depends on laser parameters. The AOS phase diagram as a function of pulse duration and fluence exhibits a characteristic triangular shape, which progressively narrows as the pulse duration increases. {\emph{Put also that the concentration may be not homogeneous and then may start in the regions with different Gd concentrations. Also there are results indicating that in multilayers the switching may start close to interfaces. }}
%{\emph{Speak about the role of the damping and that it may be different in Gd and Co.}}  
Taken together, theoretical studies consistently identify proximity to the magnetization compensation point and element-specific damping as two key ingredients for AO-HIS in Gd-based perpendicular alloys. However, recent experimental findings have challenged the former requirement. Specifically, DP Gweha Nyoma et al. \cite{3percentGdExperiment2024} demonstrated robust AO-HIS in $\text{Pt/Co/Pt}$ stacks through interfacial Gd incorporation, referred to as "Gd dusting" to emphasize the extremely low amount of rare-earth involved. Their experiments considered two wedged structures, $\text{Pt/Gd/Co/Pt}$ and $\text{Pt/Co/Gd/Pt}$, with a fixed Co thickness of $t_{\text{Co}}=0.7$ nm. Although both configurations exhibited AO-HIS, Gd incorporation at the bottom Pt/Co interface proved particularly effective, enabling switching for Gd thicknesses as low as $t_{\text{Gd}}=0.05$ nm. At this $t_{\text{Gd}}$, high-resolution transmission electron microscopy (HRTEM) could not resolve a distinct Gd layer, whereas energy-dispersive X-ray spectroscopy (EDXS) confirmed its interfacial localization. As the authors argue, a sharp Gd/Co interface does not exist and the interfacial composition is largely non-uniform. Treating the bottom interface as an effective GdCo alloy, AO-HIS is obtained at an ultra-low Gd concentration of $3\%$, placing the system very far from the magnetization compensation point and thereby challenging its presumed necessity for AO-HIS. 
\newline\indent
In this article we present results of ASD in support of the experimental work in Ref. \cite{3percentGdExperiment2024} confirming AO-HIS in a prototype Gd–Co ferrimagnet far below the magnetisation compensation point in a similar ultra-thin film geometry discussed by DP Gweha Nyoma et al. In addition, our calculations also predict AO-HIS at ultra-low Gd concentrations in an X-multilayered (XML) system where X stands for the layer repetition. The necessary condition for switching in either cases is a smaller atomistic damping constant for Gd relative to Co, consistent with previous reports considering rare-earth impurities as slow relaxing agents. \cite{Ceballos2021,Jakobs2021} 
In the following sections we detail the ASD model of the two magnetic systems we study and give a brief characterisation of their magnetic properties at thermal equilibrium. Next, we focus our attention on the ultra-fast magnetization dynamics highlighting the required conditions for AOS at ultra-low Gd concentrations as a function of the laser pulse fluence or duration and relative damping constants of Gd and Co. Interestingly, our extensive parameter sweep reveals AO-HIS can be achieved down to 1.5 \% total Gd concentration \textemdash a regime much closer to a pure ferromagnetic system and very far away from the compensation point $T_\text{M}$. In addition, we recover the characteristic triangular-shaped switching diagrams in Gd-based ferrimagnetic alloys observing AOS for a wide fs-ps range in good agreement with experimental observations. \cite{Wei2021,ZhangGdStateDiagram,3percentGdExperiment2024} Lastly, we highlight a distinctive dynamic behaviour in between the two investigated geometries wherein the XML system is faster both at short fs and longer tens of ps time-scales, yet the distinction diminishes at ultra-low Gd concentrations as the dynamics slow down in both samples. 

%\clearpage
\section{\label{sec:level1}Computational model}
The ASD calculations were carried out using the open-source software package VAMPIRE. \cite{Evans2014AtomisticNanomaterials} Figure \ref{Figure_1} shows static properties for the two material geometries we discuss: an ultra-thin film system consisting of three atomic monolayers (3ML) and a multilayered cubic structure respectively denoted hereon XML. The 3ML geometry $\left(100\times100\times 0.531\right)$ nm$^3$ stacks three Co atomic planes, where the top layer is doped with small amounts of Gd. In the XML alloy configuration $\left(17\times17\times17\right)$ nm$^3$,
\begin{figure*}
\centering
\includegraphics[width=0.95\linewidth]{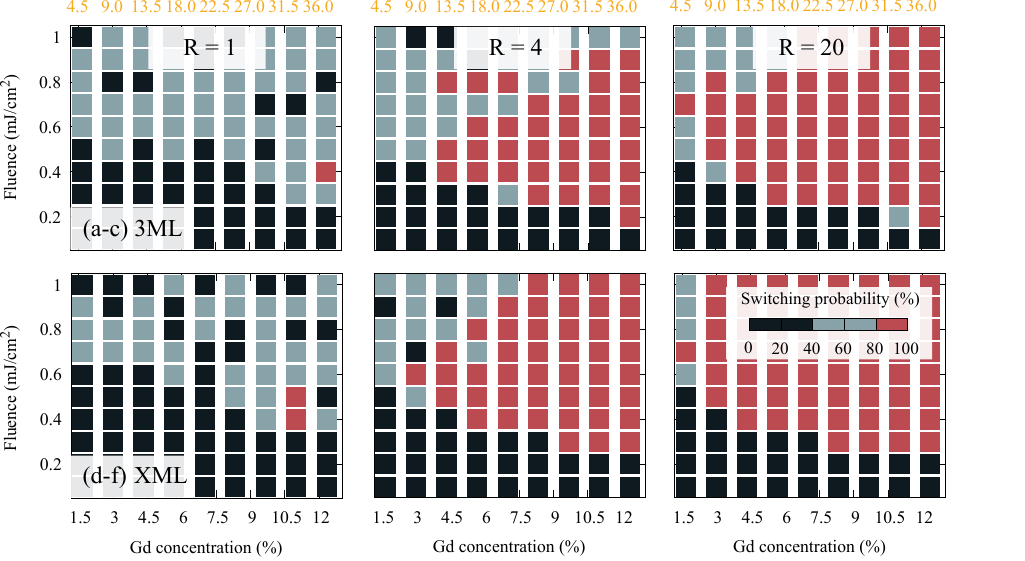}% Here is how to import EPS art
 \caption{\textbf{AOS diagrams with respect to laser fluence and rare-earth concentration:} 3ML (top row) and XML (bottom row) GdCo alloy. The pulse duration is fixed to $50$ fs. In (a-f), the bottom $x-$axis reports the Gd concentration with respect to the total atomic population. In (a-c) the top $x-$ axis denotes the Gd concentration only with respect to the atomic population in the top Co layer. From left to right, we vary the damping ratio $R=\lambda_{\text{Co}}/\lambda_{\text{Gd}}$. All subplots share the color legend denoting the switching probability $P$. The latter is calculated as the average result of ten stochastic repetitions: $P=0.1\sum_ip_i$ where $p_i$ is $+1$ for switching and $0$ for back-switching.
 }
\label{Figure_2}
\end{figure*}
we randomly distribute Gd atoms within the entire volume of the Co cubic crystal. The simulated dimensions are chosen such that the total number of atoms is in the vicinity of half a million. We assume an fcc lattice arrangement of atoms with spacing $a=3.54$ \r{A}. 
% \textcolor{red}{For the 3ML system, I have PBC in x and y. For XML, I dont have any PBC.}
The extended Heisenberg Hamiltonian
describing the competing atomic interactions is defined here as:
\begin{align}
\mathcal{H}=&-\frac{1}{2}\sum_{\langle i,j\rangle}J_{ij}\sm_i\cdot\sm_j -k_{\text{uni}}\sum_i(\sm_i\cdot\bm{z})^2,
\label{Hamiltonian}
\end{align}
where $\sm_{i,j}$ are adimensional unit vectors characterizing the orientation of the local spin magnetic moments at sites $i$ and $j$. The exchange constant $J_{ij}$ collectively denotes the three interactions present in our calculations, two of FM kind ($J_{\text{Co-Co}}$, $J_{\text{Gd-Gd}}$) and a third AFM-like exchange $J_{\text{Co-Gd}}$. To ensure perpendicular magnetic ordering at equilibrium we introduce an uniaxial anisotropy contribution with the easy-axis taken along $Oz$ and whose strength is governed by the constant $k_{\text{uni}}$. In similar fashion to Refs. \cite{Radu2011TransientSpins,Ostler2012UltrafastFerrimagnet,Ceballos2021,Jakobs2022FemtosecondCriteriaFerrimagnets}, $k_{\text{uni}}$ is assumed identical for both the Gd and Co atoms. 

Table \ref{Tabel_1} summarises the material parameters employed in our work.
\begin{table}[htbp]
\caption{GdCo material and TTM parameters used in our ASD simulations. The exchange, anisotropy, magnetic moments and lattice constants are extracted from Refs. \cite{Radu2011TransientSpins, Ostler2012UltrafastFerrimagnet,Barker2013TwoMagnon,Asmann2019SpinLatticeGilbert}. The damping constants are extracted from Refs. \cite{Barker2013TwoMagnon,Ceballos2021} and the TTM parameters are extracted from Ref. \cite{AtxitiaOstler2018_Double_pump}.}
\label{Tabel_1}
\begin{ruledtabular}
\begin{tabular}{ccc}
Parameter & Value & Unit
\\
$J_{\text{Co-Co}}$~~~$J_{\text{Gd-Gd}}$~~~$J_{\text{Co-Gd}}$& $4.35$~~~$1.26$~~~$-1.09$ & J/link$\left(\times 10^{-21}\right)$\\
$\mu_\text{s}^{\text{Co}}$~~$\mu_\text{s}^{\text{Gd}}$&$1.6$~~~$7.63$~~&$\mu_\text{B}$\\
$k_{\text{uni}}$&$8.072$&J/atom$\left(\times 10^{-24}\right)$\\
$\lambda_{\text{Co}}$~~~$\lambda_{\text{Gd}}$&$0.02$~~~0.02\textemdash0.001&adim.\\
a&$3.54$&\r{A}\\
$\gamma_{\text{e}}$&$2.25$&Jm$^{-3}$K$^{-2}$$\left(\times10^2\right)$\\
$C_{\text{ph}}$&$3.1$&Jm$^{-3}$K$^{-1}$ $\left(\times10^6\right)$\\
$G_{\text{ep}}$&$2.5$&Wm$^{-3}$K$^{-1}$$\left(\times10^{17}\right)$\\
$\tau_{\text{cool}}$&$20$&ps
\end{tabular}
\end{ruledtabular}
\end{table}
\newline\indent      
To model the ultra-fast magnetisation dynamics we integrate the set of atomistic Landau-Lifshitz-Gilbert (LLG) equations:
\begin{equation}
\frac{d\sm_i}{dt}=-\frac{\gamma}{1+\lambda_i^2}\sm_i\times\left[\mathbf{H}_{\text{eff}}^{i}+\lambda_i\left(\sm_i\times\mathbf{H}_{\text{eff}}^{i}\right)\right],
\end{equation}
where $\gamma$ is the electron gyromagnetic ratio and $\lambda_i$ is an adimensional, phenomenological intrinsic damping (coupling to the bath) parameter. The effective field  acting at each atomic site $i$ is given by: $\bm{H}_{\text{eff}}^i=-\frac{1}{\mu_0\mu_\text{s}^i}\frac{\delta\mathcal H}{\delta\sm_i}$. Here $\mu^i_\text{s}$ is the atomic magnetic moment and $\mu_0$ represents the vacuum permeability. 
\newline\indent
Starting from thermodynamic equilibrium at $300$ K, we consider a spatially uniform laser-pulse heating of the sample on the fs time-scale, which is described by the following two-temperature model (TTM) equations \cite{Kaganov1957RelaxationLattice,Lifshits1960OnRadiation,Anisimov1974ElectronPulses}:

\begin{align}
\label{electron}
C_\text{e} \frac{dT_\text{e}}{dt}&=-G_{\text{ep}}(T_\text{e}-T_{\text{ph}})+S(t)-C_\text{e}\frac{T_\text{e}-T_{\text{room}}}{\tau_{\text{cool}}},\\
C_{\text{ph}} \frac{dT_{\text{ph}}}{dt}&=G_{\text{ep}}(T_\text{e}-T_{\text{ph}}).
\label{phonon}
\end{align}
We assume the TTM parameters are identical for Gd and Co in similar fashion to Refs \cite{Radu2011TransientSpins,Ostler2012UltrafastFerrimagnet,Barker2013TwoMagnon,AtxitiaOstler2018_Double_pump,Ceballos2021,Jakobs2022FemtosecondCriteriaFerrimagnets}. Their numerical values can be found in Table \ref{Tabel_1}. $C_\text{e}$ and $C_{\text{ph}}$ are the electron and phonon volumetric heat capacities and $G_{\text{ep}}$ is a measure of the coupling between the two baths. In the free electron approximation, $C_\text{e}$ is taken as $C_\text{e}=\gamma_\text{e}T_\text{e}$. Assuming $T_{\text{ph}}$ is larger than the Debye temperature, $C_{\text{ph}}$ will be considered constant. We introduce also a simple Newton cooling law to model the heat transfer with the external medium ($T_{\text{room}}=300$ K) at a rate given by $\tau_{\text{cool}}$. The time dependent laser pulse power is given by \cite{Jiang2005TTM}:
\begin{equation}
S(t)=\frac{2F}{t_\text{p}\delta\sqrt{(\pi/\ln2)}}\exp\left[-4\ln2\left(\frac{t-3t_\text{p}}{t_\text{p}}\right)^2\right],
\label{lpulse_power}
\end{equation}
where $F$ is the laser fluence at full width half maximum (FWHM), $\delta$ is the optical penetration depth and $t_\text{p}$ is the pulse duration. We assume all energy from the laser is absorbed in the sample and the penetration depth $\delta_{\text{opt}}$ is considered equal to the material's thickness. The larger-scale switching phase diagram simulations used a supercomputer to accelerate the calculations.\cite{archer2} 

\section{\label{sec:level1}Results}
In Figure \ref{Figure_1} we display the equilibrium net magnetisation \textemdash calculated via $|{\bm{m}}|=|\sum_i\mu_\text{s}^i\sm_i|/\sum_i\mu_\text{s}^i$ \textemdash as a function of temperature and at different alloy concentrations in the XML and 3ML configuration. Reported with respect to the total atomic population, we limit the Gd concentration to a maximum of $x_{\text{Gd}}=13\%$. As expected, the reduced $Oz$ dimension of the ultra-thin film geometry diminishes overall the critical Curie temperature compared to the XML system. At $x_{\text{Gd}}=0\%$, following the fitting procedure discussed in Refs. \cite{KazantsevaMultiscale2008,OstlerPRB2011Crystalographically} which neglects finite-size effects in the $m(T)$ curve, we obtain $T_\text{c}^{\text{3ML}} \approx 612$ K and $T_\text{c}^{\text{XML}}\approx 987$ K respectively. In the rare-earth concentration range we simulate here,  no $T_\text{M}$ point appears, consistent with the experiments and calculations in Ref. \cite{OstlerPRB2011Crystalographically}, where in XML for instance, $T_\text{M}$ becomes apparent in the vicinity of $x_{\text{Gd}}=24\%$. Here, the contribution to the net magnetisation originating from the transition metal remains dominant and therefore the critical temperature of the alloy is largely governed by Co.
\newline\indent
\begin{figure}
\centering
\includegraphics[width=1.0\linewidth]{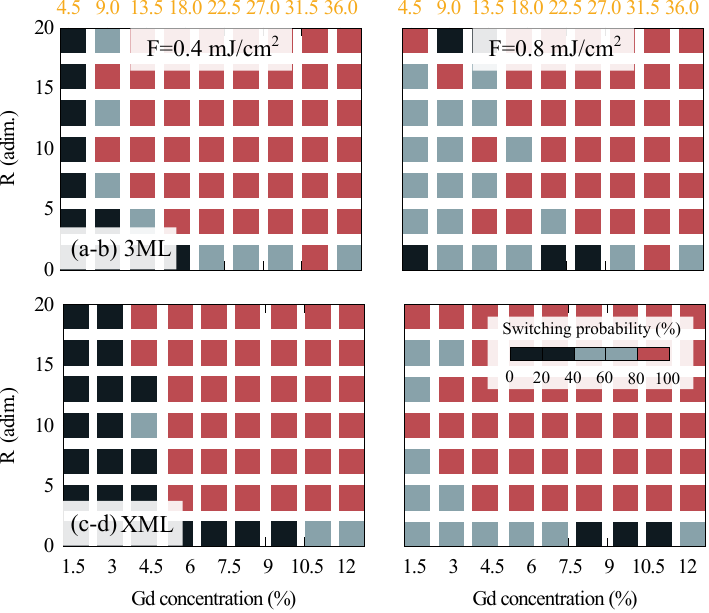}\caption{\textbf{AOS diagrams with respect to rare earth concentration and damping ratio:} 3ML (top row) and XML (bottom row) GdCo alloy. From left to right we vary the laser pulse fluence. The yellow highlighted percentage represents the Gd concentration calculated with respect to the top Co layer in the 3ML configuration.}
\label{Figure_3}
\end{figure}
Figure \ref{Figure_2} presents AOS diagrams as a function of laser fluence and Gd concentration at three distinct damping ratios $R=\lambda_{\text{Co}}/\lambda_{\text{Gd}}$. Every switching probability point $P$ in the diagrams is calculated as the average of N stochastic repetitions: $P=\frac{1}{N}\sum_ip_i$ where $p_i$ is $+1$ for switching and $0$ for back- or non- switching. Due to very large system size typically $N=10$ was used.  The laser pulse duration is fixed at $50$ fs. Furthermore, we fix $\lambda_\text{Co}$ at $0.02$ and decrease $\lambda_\text{Gd}$ with respect to this threshold. 
As observed in subplots (a) and (d), an identical damping in the Co sublattice and Gd ($R=1$) results in random thermal switching or no AOS. On the other hand, increasing $R$ (decreasing the damping in Gd) leads to an expansion of the red AOS area corresponding to $P>80\%$.  Revealed in subplots (c) and (f), at $R=20$ the area of most probable switching extends from the highest Gd concentration of $x_{\text{Gd}}=13\%$ we simulate, down to the surprisingly low threshold of $x_{\text{Gd}}=3\%$. Carefully tailoring the laser pulse fluence, our results show this lower limit can further be decreased to $x_{\text{Gd}}=1.5\%$ which brings our system closer to a pure ferromagnetic state rather than a ferrimagnet. In addition, a decrease of the threshold fluence to switching can be observed with respect to increasing the Gd concentration at $R=4$ and $R=20$. Interestingly, an upper fluence threshold for which AOS occurs at a constant low rare-earth concentration can be identified particularly in subplots (b), (c) and (e) of the same Figure \ref{Figure_2}. This upper threshold does not appear in the XML case for considered fluencies at $R=20$ as seen in (f), constituting here the only qualitative distinction with respect to 3ML configuration. 
\newline\indent
\begin{figure*}
\centering
\includegraphics{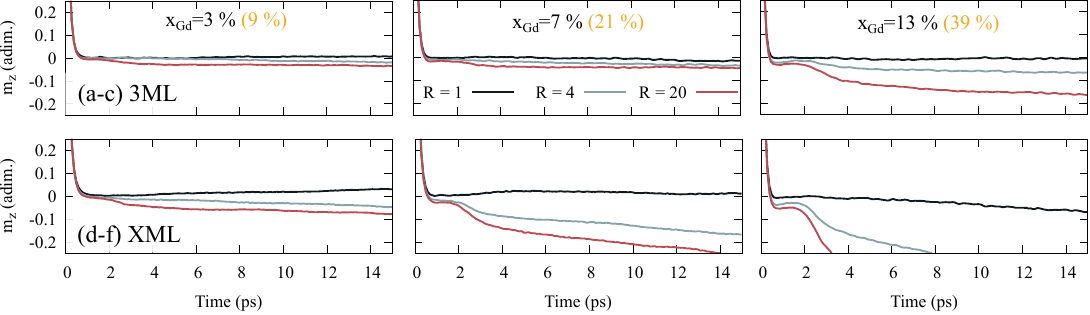}% Here is how to import EPS art
\caption{\textbf{Perpendicular AOS magnetisation dynamics with respect to damping and RE concentration:} 3ML (top row) and XML (bottom row) GdCo alloy. The fluence and pulse duration are fixed at $0.55$ mJ/cm$^2$ and 50 fs respectively. From left to right we vary the RE concentration $x_{\text{Gd}}$ calculated with respect to the total atomic population. The yellow highlighted percentage represents the Gd concentration calculated with respect to the top Co layer in the 3ML configuration. Each subplot displays AOS dynamics at three distinct damping ratios: $R=1,4$ and $20$ respectively. The $m_z$ magnetisation component represents the net contribution involving both the Gd and Co phase which we calculate via: $m_z=\sum_i\mu_iS_i^z/\sum_i\mu_i$.}
\label{Figure_4}
\end{figure*}
In Figure \ref{Figure_3} we present AOS diagrams as a function of the damping ratio $R$ and the concentration $x_{\text{Gd}}$ once again for both alloy geometries. While the laser pulse duration is fixed at $50$ fs for all investigated cases, we calculate AOS diagrams at two distinct laser fluences: $0.4$ and $0.8$ mJ/cm$^2$. From these results it is clear that increasing $R$ or in other words decreasing the Gd damping value shifts the minimum $x_{\text{Gd}}$ concentration for which switching occurs to lower values. For instance, both in the XML and 3ML case, at a fluence of $0.8$ mJ/cm$^2$ and a damping 
ratio of $R=19$ one can obtain AOS for rare-earth concentration as small as $x_{\text{Gd}}=1.5$ \% \textemdash seen in subplots (b) and (d) of Figure \ref{Figure_3}.
\begin{figure*}
\centering
\includegraphics{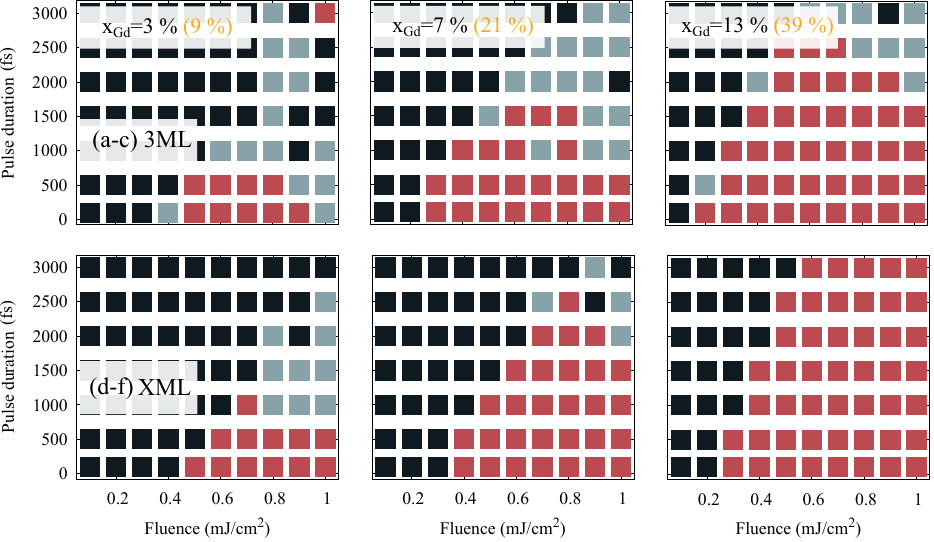}% Here is how to import EPS art
\caption{\textbf{AOS diagrams with respect to laser fluence and pulse duration:} 3ML (top row) and XML (bottom row) GdCo alloy. From left to right we vary the RE concentration $x_{\text{Gd}}$. Here, the damping ratio $R$ is fixed at $20$. The yellow highlighted percentage represents the Gd concentration calculated with respect to the top Co layer in the 3ML configuration.}
\label{Figure_5}
\end{figure*} 
\begin{figure}
\centering
\includegraphics[width=0.9\linewidth]{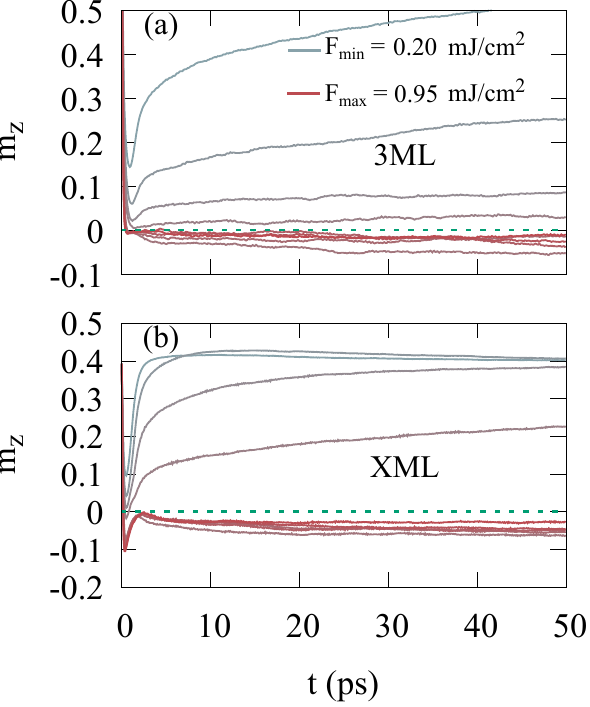}% Here is how to import EPS art
\caption{\textbf{Ultra-fast magnetisation dynamics vs laser fluence in:} (a) 3ML and (b) XML alloy configuration. The pulse duration is fixed at $50$ fs, the damping ratio is $R=20$ and the Gd concentration is $x_{\text{Gd}}=3 \%$.}
\label{Figure_6}
\end{figure} 
\newline\indent
In addition, the magnetisation dynamics reveal very interesting details in the AOS process of the two alloy geometries. Figure \ref{Figure_4} showcases examples of the ultra-fast perpendicular magnetisation dynamics displayed as a function of $R$ and $x_{\text{Gd}}$. The fluence and pulse duration are fixed at $0.55$ mJ/cm$^2$ and 50 fs respectively. We focus our attention on the net $m_z$ magnetisation component including both the Gd and Co contribution which we calculate via: $m_z=\sum_i\mu_\text{s}^iS_i^z/\sum_i\mu_\text{s}^i$. In all the switching events reported in Figure \ref{Figure_4} we observe a two-step process characteristic of GdCo alloys: this AOS takes place via an initial fast demagnetisation on a time-scale of hundreds of fs followed by a subsequent slow recovery of the magnetisation on a time-scale above tens of ps. Interestingly, in subplots (a) and (d) corresponding to $x_{\text{Gd}}=3$ \%, we identify switching dynamics resembling the result reported in Figure 3 of Ref. \cite{3percentGdExperiment2024}. Although their Pt/Co/Pt multilayer with Gd dusting at the bottom Pt/Co interface rather resembles the 3ML alloy configuration here, our results suggest AOS very far away from magnetisation compensation could potentially be achieved also in the XML system. Increasing the Gd concentration to $7$ \% and $13$ \% respectively leads to faster AOS dynamics in both system geometries seen in subplots (b,c) and (e,f). 
%This is in agreement with the predictions of Ref.\cite{Suarez} based on mean-field theory that indicates that the dynamics of both Gd and Fe sublattices accelerate as a function of Gd concentration at high temperatures.
%the longitudinal magnetisation relaxation timescale in alloys scale is inversely proportional to $H_{ex}^{\nu}=\mu_{\nu}( x_{\nu}J_{0,\nu}m_{\nu}+(1-x_{\nu}|J_{0,\nu \kappa})|m_{\kappa})$ where $\nu,\kappa$ stand for Gd or Fe contributions   and $J_0$ for the corresponding exchange parameters, including the average nearest number of neighbors in the undoped lattice. Considering that the system is dominated by Fe, its sublattice accelerates as a function of Gd concentration for small $x$. 
Overall however, XML GdCo switches and recovers the magnetisation much faster than the 3ML alloy configuration at a fixed $R$ ratio and $x_{\text{Gd}}$ as can be observed particularly in subplots (b-f) of Figure \ref{Figure_4}. This in the first place comes from the fact that in the 3ML alloy Gd is present only at the interface where the switching is initiated while in the XML system the switching nuclei \textemdash known as magnon drops \cite{Kosevich1990,Iacocca2019} \textemdash are distributed over the volume. Additionally the 3ML system has a lower Curie temperature (due to a lack of neighbors on the surface) also pointing to a slower magnetisation dynamics.
\newline\indent
Another interesting result which also needs to be underlined here is a counter-intuitive speedup of the AOS dynamics via an increase of the damping ratio $R$, observed in all investigated scenarios. An unitary $R$ on the other hand leads to demagnetisation or random switching events. This matches the conclusion in Ref. \cite{Ceballos2021} where ASD simulations revealed that reducing the rare-earth damping leads to faster AOS dynamics in Gd alloys. Finally, the stochastic nature of this switching problem highlights also several back-switching events in Figure \ref{Figure_4} observed for instance in subplots (a), (d) and (e).
\newline\indent
The AO-HIS in Gd-based ferrimagnets follows a well-known, triangular laser fluence vs pulse duration state diagram discussed previously for instance in Refs. \cite{Wei2021,ZhangGdStateDiagram}. Remarkably, the AOS we obtain in the absence of $T_{\text{M}}$ also displays this type of diagram as exemplified in Figure \ref{Figure_5}. It is important to underscore the fact that the damping ratio $R$ is fixed here at $20$. In contrast to the results reported in Ref. \cite{3percentGdExperiment2024} our state diagrams obtained at $x_{\text{Gd}}=3$ \%, either in the XML or 3ML case, showcase a smaller limiting pulse duration for which AOS can still occur, fixed below $1$ ps \textemdash as seen in subplots (a) and (d). This threshold is overcome increasing the $x_{\text{Gd}}$ concentration to $7$ \% and $13$ \% where the AOS takes place up to a maximum of $2-3$ ps, a limiting value consistent with the results in Ref.\cite{3percentGdExperiment2024}. Nonetheless, the triangular shaped state diagram marking the AOS region is qualitatively recovered for all investigated $x_{\text{Gd}}$ values. In general, our results indicate the switching obtained in the XML configuration is more robust against the variations of the laser pulse parameters as demonstrated by subplots (e) and (f) contrasted with the 3ML results in (b) and (c) respectively. 
This distinction can stem from the non-equivalent critical temperatures of the two alloys which naturally sets a lower limit for the maximum absorbed laser power in the 3ML system. 
\newline\indent
It is also important to note here that within our ASD calculations it is challenging to set the boundary between a multi-domain state and a switched configuration at large laser pulse fluences due to the absence of the demagnetising term in the total effective field and the slow nature of this AOS process particularly in the magnetisation recovery region. At a time-scale of ns, domain formation and domain-wall dynamics  should complete the reversal process. In our simulations, however, large timescale is not achievable and the system size is too small to allow formation of domains and their expansion governed by long-range dipolar interactions. In our simulations, 
the net magnetisation does not fully recover  hundreds of ps after switching (see supplementary material Note S1)
%\textcolor{red}{[This is temporary data. I am running some new sims.]}. 
Nonetheless, 
the ultra-fast magnetisation dynamics we obtain on the fs-ps time scale qualitatively reproduce the behaviour observed experimentally in Ref. \cite{3percentGdExperiment2024}. 
Although in our simulations the maximum electron temperature $T_{\text{e}}$ can overcome the $T_\text{c}$ point by several hundreds of K, $T_{\text{e}}$ relaxes back to room temperature in a matter of ps, even at the maximum fluence $F=1$ mJ/cm$^2$ we simulate (see for instance the results presented in the supplementary material Note S2). 
Furthermore, the issue of distinguishing between demagnetised and switched states persists at low Gd concentrations (for instance $x_{\text{Gd}}=3$ \%) also in the case of low laser fluences due to the nature of the slow AOS reversal for instance observed in Figure \ref{Figure_4}(a) or discussed in Ref. \cite{3percentGdExperiment2024}. This motivates the use of the straightforward AOS test where the perpendicular magnetisation crossing the $m_z=0$ threshold is sufficient to identify a positive switching event. This test we believe holds only in the low ps time-scale. To highlight the sensitivity to the AOS criterion, we display in supplementary material Note S3 the same data in Figure \ref{Figure_2} but evaluate switching using the condition: $m_z<-0.1$.
\newline\indent
Finally, another result closely matching the experiments reported in Ref. \cite{3percentGdExperiment2024}  is presented in Figure \ref{Figure_6}. The 3ML AOS dynamics vs laser fluence $F$ given in subplot (a) qualitatively reproduce the TR-MOKE data presented in Figure 3 of Ref. \cite{3percentGdExperiment2024}. This confirms the assumption formulated in the aforementioned experimental work according to which the transition metal within the Pt/Co/Pt multilayer system with Gd dusting at the Pt/Co interface behaves in essence as a GdCo alloy. Interestingly, subplot (b) shows the AOS dynamics beyond 5 ps in the XML configuration to be in accordance with the 3ML system, but with clear distinctions below this threshold. In the XML, the initially switched magnetisation recovers temporarily towards the opposite direction before a second slower demagnetisation and AOS process occurs, a behaviour for instance exemplified in Ref \cite{Sergiu_pair}.
\section{\label{sec:level1}Conclusions}
In this work, we have performed ASD simulations to investigate AO-HIS in the prototype GdCo ferrimagnet with very low Gd concentrations, both in a XML  and ultrathin film geometries. Our results demonstrate that low intrinsic damping of the Gd sublattice is a key requirement for achieving robust single-pulse switching in a very low Gd concentration regime. However, both during switching and ultrafast-demagnetisation the XML alloys recover faster than ultrathin dusted films. This highlights the strong influence of interfacial effects on the relaxation pathways.
\newline\indent
A notable and counterintuitive result of our study is that increasing the Gd concentration and reducing its damping accelerates the dynamics. This behavior contrasts with expectations from the high-Gd concentration regime \cite{Koopmans2009,Atxitia2011,Atxitia2014}, where the dynamics are typically described by the scaling criterion $\mu/\lambda T_\text{C}$, predicting slower dynamics with increasing rare-earth contribution and decreasing its damping. Our findings therefore suggest that this commonly used ultrafast demagnetisation criterion is not applicable in the dilute Gd limit.
Instead, we propose an alternative physical picture governing the low-concentration regime. When the Gd content is very small, the switching process likely proceeds via the formation of a limited number of localized reversal regions (magnon-droplet-like excitations \cite{Iacocca2019}). These regions are highly fluctuating in temperature and must survive, expand and merge to complete the magnetisation reversal. At low Gd concentration and damping, the  probability of survival time of such regions is strongly reduced, leading to fewer reversal centers and consequently slower global recovery dynamics. This mechanism provides a natural explanation for the observed slowdown in magnetisation dynamics at low Gd concentration.
% \textcolor{red}{Paul: I think the claims above although previously demonstrated in literature, here they are not well supported, maybe we should clearly show in a figure the formation of these drops as a function of time like in Pablo's paper or Ref \cite{Iacocca2019}}
Overall, our results demonstrate that proximity to the magnetisation compensation point is not a prerequisite for AO-HIS. Instead, efficient switching can be achieved with ultralow Gd concentrations, provided that the intrinsic damping remains sufficiently low. This finding has important implications for materials design and technological applications, as it enables the reduction of rare-earth content while maintaining functional switching behavior.

\begin{acknowledgments}
This work made use of the VIKING cluster, provided by the University of York as well as the ARCHER2 UK National Supercomputing Service (https://www.archer2.ac.uk). The authors acknowledge funding from the European Union’s Horizon 2020 research and innovation programme under the Marie Skłodowska-Curie International Training Network COMRAD (grant agreement No 861300). 
\end{acknowledgments}

\section*{Data Availability Statement}
The data that support the findings of
this study are available from the
corresponding author upon reasonable
request.

\nocite{*}
\bibliography{References_byhand}% Produces the bibliography via BibTeX.

\end{document}

%\preprint{APS/123-QED}

\title{Atomistic Insights into the All-Optical Switching of GdCo Alloys with Minimal Gd Concentration} % Force line breaks with \\
% \thanks{A footnote to the article title}%

\author{Paul-Iulian Gavriloaea}
\affiliation{National Institute of Research and Development for Technical Physics, 47 Mangeron Blvd., 700050 Iasi, Romania}
\affiliation{Instituto de Ciencia de Materiales de Madrid, CSIC, Cantoblanco, 28049 Madrid, Spain}
%Lines break automatically or can be forced with \\
\author{Jackson L. Ross}%
\affiliation{ 
Institute for Condensed Matter Physics and Complex Systems, The
University of Edinburgh, Edinburgh, EH9 3FD, United Kingdom}%
\affiliation{School of Physics, Engineering and Technology, University of York, YO10 5DD, York, UK}
\author{Richard F. L. Evans}
\affiliation{School of Physics, Engineering and Technology, University of York, YO10 5DD, York, UK}
\author{Roy Chantrell}
\affiliation{School of Physics, Engineering and Technology, University of York, YO10 5DD, York, UK}
\author{Stéphane Mangin}
\affiliation{Université de Lorraine, CNRS, IJL, Nancy, France}
\author{Oksana Chubykalo-Fesenko}
\affiliation{Instituto de Ciencia de Materiales de Madrid, CSIC, Cantoblanco, 28049 Madrid, Spain}

%\keywords{Suggested keywords}%Use showkeys class option if keyword
                              %display desired
\maketitle

%\tableofcontents
\onecolumngrid
\newpage

\section*{Note S1. }
\begin{suppfigure}[!h]
    \centering
    \includegraphics[width=\textwidth]{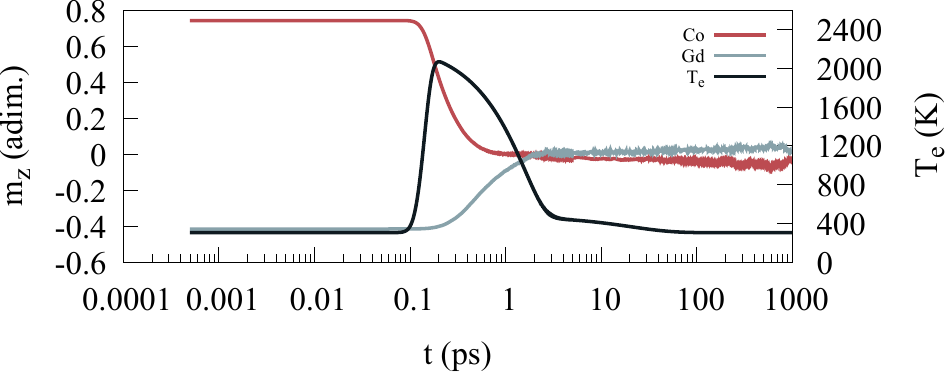}    \caption{\textbf{Example of slow magnetisation recovery process after AOS in the 3ML system.} The Gd concentration (with respect to the whole atomic volume) is $x_{\text{Gd}}=3$ \%, the laser pulse duration and fluence are $t_{\text{p}}=50$ fs and $F=50$ mJ/cm$^2$ respectively, while the damping ratio is set to $D_\text{R}=20$.}
    \label{fig:S1}
\end{suppfigure}
The AOS dynamics we simulate in the low-Gd concentration limit display a very slow magnetization recovery process which extends far into the ps-ns regime thus exceending our computational capabilities. An example of this slow recovery process can be observed in Figure S\ref{fig:S1} where the Co and Gd $m_z$ magnetisation components are displayed in conjunction with the electron temperature dynamics $T_\text{e}$. On ps to ns timescales, the reversal is expected to involve domain nucleation and domain-wall propagation, processes largely governed by long-range dipolar interactions that are not accounted for in our calculations. Furthermore, the limited simulation times and relatively small system dimensions prevent us from capturing the full development of these processes. Nonetheless, the AOS dynamics we obtain in the short fs-ps time-scale matches the TR-MOKE data in Ref. \cite{3percentGdExperiment2024}.

\section*{Note S2.}
\begin{suppfigure}[!h]
    \centering    \includegraphics[width=\textwidth]{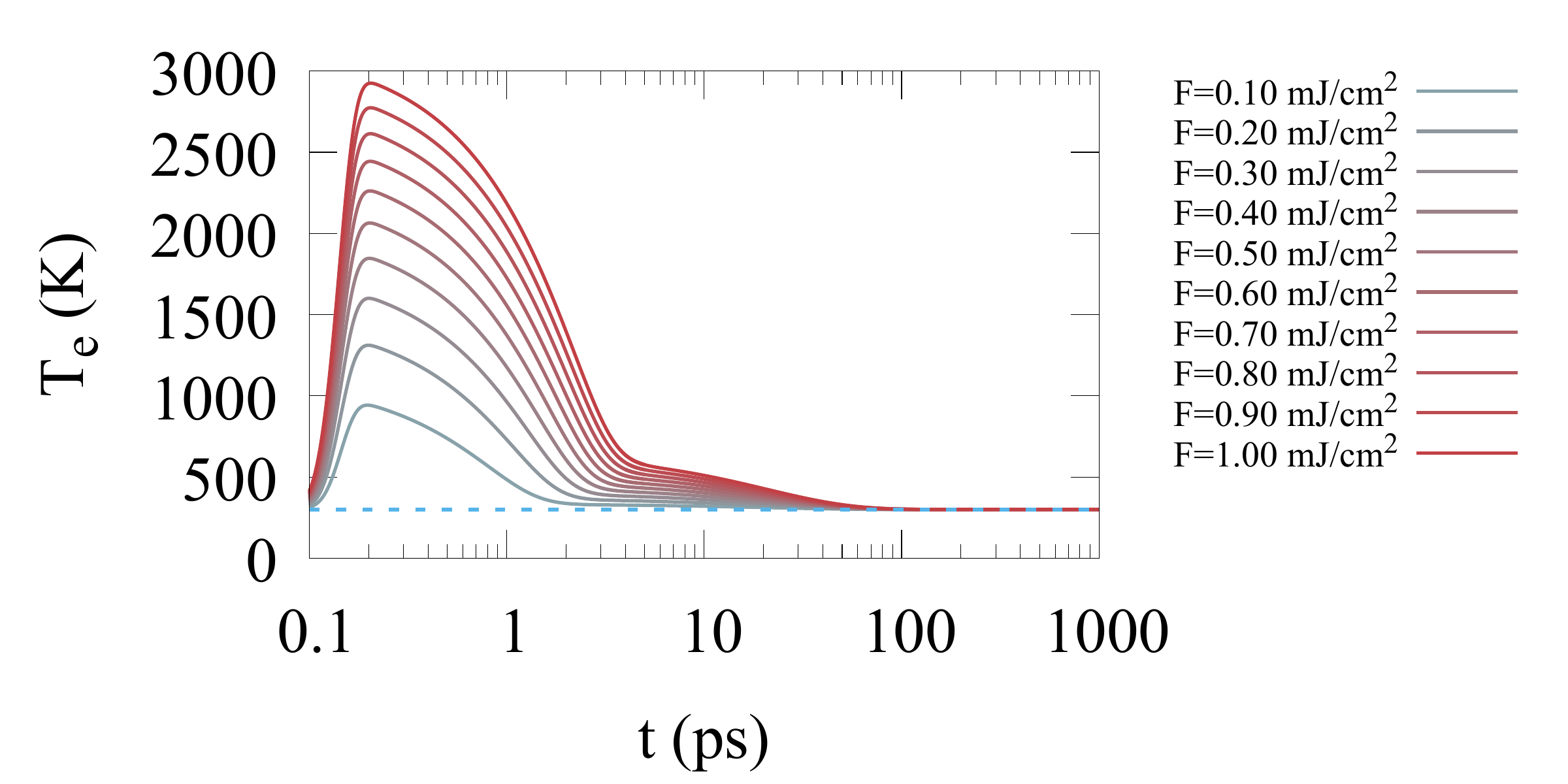}
    \caption{\textbf{Ultra-fast laser pulse heating of the electron bath as a function of fluence.} The x-axis is given in logarithmic scale. The blue dashed line corresponds to a temperature of $300$ K. The corresponding material parameters and TTM equations can be consulted in the main text.}
    \label{fig:S2}
\end{suppfigure}
Distinguishing a genuinely switched magnetic state from a randomly demagnetised configuration becomes particularly challenging in the low-Gd concentration regime, owing to the slow magnetisation recovery characteristic of these systems. An important consideration is therefore to ensure that the electron temperature does not remain above the Curie temperature over extended picosecond-to-nanosecond timescales, as this could favour a demagnetised or disordered magnetic state rather than a stable reversal.

As it can be seen in Figure S\ref{fig:S2}, the maximum electron temperature $T_{\text{e}}$ in our simulations exceeds $T_{\text{C}}$ by several hundred kelvin but it relaxes back to room temperature within a few to 10s of picoseconds, even at the highest simulated fluence of $F=1$ mJ/cm$^2$. This indicates that the electron-temperature dynamics are unlikely to be responsible for a persistent demagnetised state, as the system rapidly cools below $T_{\text{C}}$.

\section*{Note S3.}

\begin{suppfigure*}[!h]
    \centering
\includegraphics[width=0.93\linewidth]{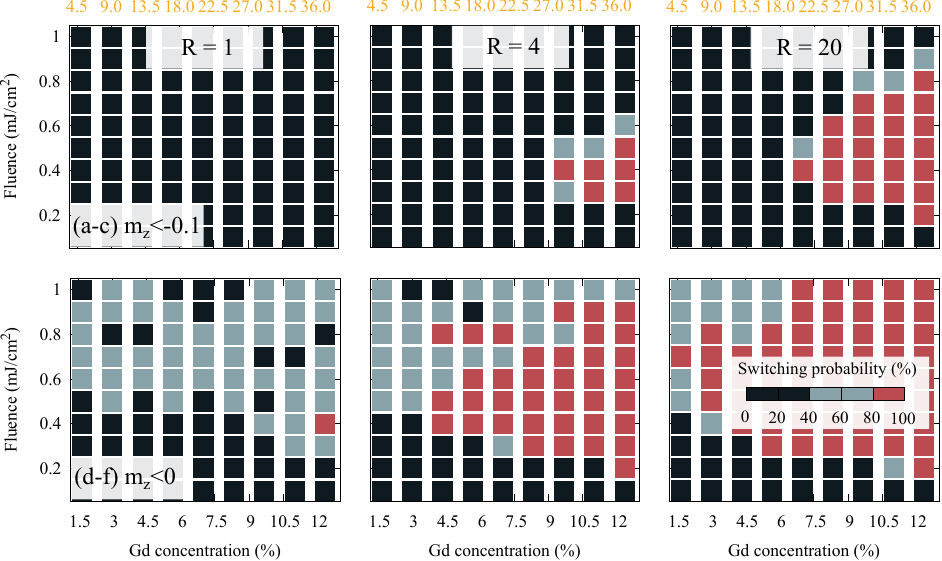}
    \caption{\textbf{Switching criterion comparison in constructing 3ML AOS diagrams:} (a-c) $m_z<-0.1$ and (d-f) $m_z<0$. The latter three subplots are also given in Figure 2 of the main text. }
    \label{fig:S3}
\end{suppfigure*}

\begin{suppfigure*}[!h]
    \centering
\includegraphics[width=0.93\linewidth]{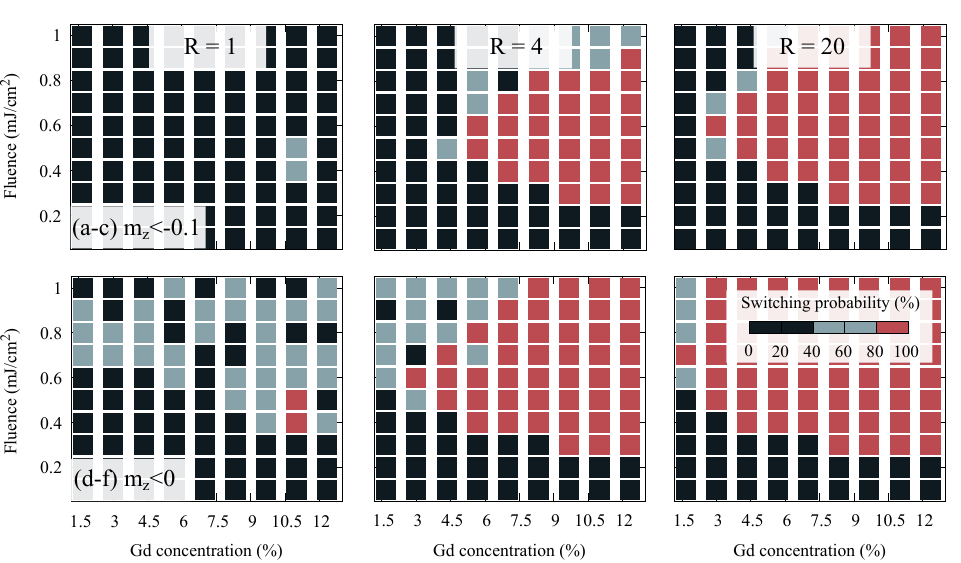}
    \caption{\textbf{Switching criterion comparison in constructing XML AOS diagrams:} (a-c) $m_z<-0.1$ and (d-f) $m_z<0$. The latter three subplots are also given in Figure 2 of the main text. }
    \label{fig:S4}
\end{suppfigure*}

An important consideration in constructing the all-optical switching (AOS) diagrams is the choice of the switching criterion. Given the inherent difficulties in unambiguously distinguishing a switched state from a demagnetised configuration, as discussed in the main text, we adopt a straightforward criterion based on the sign reversal of the out-of-plane magnetisation, i.e., $m_z<0$. Accordingly, a switching event is identified whenever the net perpendicular magnetisation crosses the $m_z=0$ threshold by the end of the simulation. We note, however, that this criterion is primarily intended to capture the short-time reversal dynamics on the picosecond timescale and does not necessarily guarantee the formation of a stable switched state at longer times. To illustrate the sensitivity of the AOS diagrams to the choice of switching criterion, Figures S\ref{fig:S3} and S\ref{fig:S4} reconstruct the 3ML and XML AOS diagrams in Figure 2 of the main text using a more stringent criterion, namely $m_z<-0.1$. Although the optimum switching area (red squares) is drastically reduced, the qualitative variation of AOS as a function of fluence, damping ratio $R$ and rare-earth concentration $x_\text{Gd}$ is preserved. 

\bibliography{References_byhand.bib}

% Produces the bibliography via BibTeX.